\documentclass[11pt]{article} 

\usepackage{amsmath}

\usepackage[adobe-utopia]{mathdesign}
\usepackage{graphicx}
\usepackage{caption}
\usepackage[inner=1.5in, top = 0.8in]{geometry}
\usepackage[hidelinks]{hyperref}
\usepackage{cleveref}

\usepackage{natbib}

\usepackage[dvipsnames, table]{xcolor}
\definecolor{maincol}{rgb}{0, 0.29, 0.33}
\definecolor{linkcol}{RGB}{0, 0, 0}
\definecolor{newforMW}{rgb}{0, 0, 1}

\usepackage{lineno}
\usepackage{paralist}

\usepackage{helvet}
\usepackage{titlesec}
\titleformat{\section}
  {\normalfont\Large\bfseries\sffamily\color{maincol}}{\thesection.}{1em}{}
\titleformat{\subsection}
  {\normalfont\large\bfseries\sffamily}{\thesubsection}{1em}{}
\titleformat{\subsubsection}
  {\normalfont\normalsize\itshape\sffamily}{\thesubsubsection.}{1em}{}

\makeatletter
    \def\@maketitle{%
  \newpage
  \null
  \vskip 2em%
  \let \footnote \thanks
   \noindent {\Large \bfseries \sffamily \color{maincol} \@title \par}%
    \vskip 1.5em%
    \noindent {\normalsize
        \@author
	\par}%
    \vskip 0.5em%
   {\scriptsize \noindent \@date \par
   }
  \par
  \vskip 1.5em}
\makeatother

\newcommand{\quotestyle}[1]{{``\emph{#1}''}}

\newcommand{\Covid}{\mbox{COVID-19}}

\usepackage{tikz}

\usepackage{CJKutf8}

\title{No evidence of systematic proximity ascertainment bias in early \Covid{} cases in Wuhan \\ {\large Reply to Weissman (2024)}}
\author{F. D\'ebarre$^1$ \& M. Worobey$^2$}
\date{}

\begin{document}

\maketitle

{\noindent \small $^1$ 
Institut d'Écologie et des Sciences de l'Environnement (IEES-Paris, UMR 7618), CNRS, Sorbonne Université, UPEC, IRD, INRAE, Paris, France. ORCID: 0000-0003-2497-833X.\\ Contact: \mbox{florence.debarre@sorbonne-universite.fr} \\
$^2$ Department of Ecology and Evolutionary Biology, University of Arizona, Tucson, AZ, USA. \\ Contact: \mbox{worobey@arizona.edu}}\\

In a short text published as Letter to the Editor of the \emph{Journal of the Royal Statistical Society Series A}, \citet{Weissman2024} argues that the finding that early COVID-19 cases without an ascertained link to Wuhan's Huanan Seafood Wholesale market resided on average closer to the market than cases epidemiologically linked to it, reveals \quotestyle{major proximity ascertainment bias}. Here we show that Weissman's conclusion is based on a flawed premise, and that there is no such \quotestyle{internal evidence} of major bias. The pattern can indeed be explained by places of infection not being limited to residential neighbourhoods, and by stochasticity – i.e., without requiring any ascertainment bias. 

After digitising and analysing data shared in the report of the 2021 Joint WHO-China Study on Origins of SARS-CoV-2, \citet{Worobey2022} showed that residential locations of early \Covid{} cases in Wuhan were strikingly distributed around the Huanan market. Given the role played by the Huanan market in the initial discovery of the outbreak \citep{Worobey2021, Yang2024book}, it was important to ensure that the spatial pattern observed was not caused by ascertainment bias. If present and major, ascertainment bias would indeed preclude any inference of a potential source of the outbreak. \citet{Worobey2022} therefore checked the robustness of their conclusions to heterogeneities in case ascertainment (see their Fig.~S12). By gradually removing the cases residing closest to the market, \citet{Worobey2022} found that substantial fractions of cases had to be removed to lose significance, demonstrating that the overall centrality of the Huanan market among early cases was not artefactually driven by cases residing the closest to the market.

\Citet{Weissman2024} however claimed to have discovered \quotestyle{internal evidence} of ascertainment bias. With the available data \citep{WHO2021}, \citet{Worobey2022} found that cases epidemiologically unlinked to the Huanan market resided on average closer to the Huanan market than cases epidemiologically linked to it. According to \citet{Weissman2024}, this difference is internal evidence of bias, because (in the absence of any proffered model to explain why) he expected unlinked cases to be further away from the initial source than linked cases.

This expectation, however, is incorrect: there is no reason the residential locations of early unlinked cases should necessarily be, on average, further away from the market than the residential locations of linked cases. Instead, given the history of the outbreak, there was ample reason to expect they would be closer: as noted by \citet{Worobey2022}, the residential locations of linked cases (primarily, individuals who worked at the Huanan market) simply reflects their average commuting distance from their workplace: \quotestyle{For market workers, the exposure risk was their place of work, not their residential locations} \citep{Worobey2022}. For unlinked cases, on the other hand, one of their exposure risks was living close to the market itself and thereby being exposed to infected market-linked cases. Such exposure risks were documented, for example in clinics very near the market where infected Huanan market workers sought care.\footnote{\begin{CJK*}{UTF8}{bsmi} 寻找华南海鲜市场第一个感染者\end{CJK*} (Searching for the first infected person in Huanan Seafood Market), March 2020, \url{https://archive.is/iPJ1x}} Moreover, many infections with SARS-CoV-2 are asymptomatic or mild, and transmission can occur prior to symptom onset \citep{Buitrago2022asymp}. Thus, to whatever extent market-linked infected individuals shared space with locals at nearby shops, restaurants, bars, and other locations, the earliest market-unlinked cases would be expected to reside nearer the market, as long as the average commuting distance of market-linked individuals was substantial (and it was). Weissman's argument implicitly assumes that individuals who worked at or visited the market were hermetically sealed off from people who lived in the neighbourhoods surrounding the market, and themselves lived very near the market. These assumptions are refuted by evidence. 

Weissman's reasoning conflates residential locations and places of infection. His (unstated, implicit) model erroneously assumes that the distributions of distances between where cases were infected and where they resided are the same for market-linked and market-unlinked cases. The locations reported in the maps of \Covid{} cases with onset in December 2019 \citep{WHO2021} and extracted by \citet{Worobey2021, Worobey2022} are residential locations. While household transmission plays an important part in the spread of \Covid{} \citep{Li2021Household}, the Huanan market was identified as place of exposure for most cases with epidemiological links to the market; it was clearly the place of infection for individuals in established transmission chains in the market \citep{Li2020MarketCluster}. These cases were demonstrably infected in the Huanan market, but the reported location in the case data is their residence, which in some cases was very far from the market \citep{WHO2021}. Secondary infections were then more likely to occur near locations with high densities of infected individuals, i.e. in the Huanan market's neighbourhood. If it were possible to know all places of infection (rather than residence), the case map would look different; there would be a large number of points on the Huanan market, corresponding to market-linked cases. Instead, the distances reported for market-linked cases who worked at the market are commuting distances, and are therefore likely to be greater than those corresponding to infections of nearby residents. 

This difference in typical spatial distances also plays a role in an additional explanation for why early unlinked cases could live closer, on average, to the market than early market linked cases, even under the unrealistic assumption that all reported case residential locations (including those of market workers) correspond to places of infection. Even if that were indeed the case, then early secondary cases could still live, on average, closer to the initial source than primary cases. \Citeauthor{Weissman2024}'s argument is described in terms of average outcomes (e.g., \quotestyle{The mean-square displacement (MSD) from the HSM of the unlinked cases would then be approximately the sum of the MSD of the linked cases and the MSD from the linked cases of the remaining steps in which traceability was lost.}) -- neglecting the fact that the number of cases is not infinite, nor large enough to be considered as infinite. Assuming isotropy, i.e., that a secondary case can reside at any angle from their infector's residence, then, by chance, a secondary case can obviously reside closer to the market than their infector. Consider a market-linked infection residing at distance $\rho_0$ from the market, and their infectee residing at distance $\rho_1$ from the infector: if the infectee can reside at any angle from the residence of their infector (i.e., assuming isotropy), then the probability that the infectee resides closer to the market than the infector is given by:
\begin{displaymath}
q = \frac{\cos^{-1}\left(\frac{\rho_1}{2 \rho_0}\right)}{\pi} . 
\end{displaymath}
This probability is equal to $1/3$ when $\rho_1 = \rho_0$, and increases to $1/2$ when $\rho_1 \to 0$. Contrary to the central premise of Weissman's argument, and in light of the fact that the dataset is not just finite but quite small, the unlinked cases will not necessarily reside on average farther away from the Huanan market than the linked cases. 

To illustrate one aspect of this point, we simulated an infection process, as follows. \begin{inparaenum}[(1)]\item We first draw $n_0$ market-linked infections and their spatial positions, at average distance $r_0$ from the market (the market is positioned, without loss of generality, at the origin); distances are drawn from an exponential distribution. \item We then draw numbers of secondary infections (i.e., market unlinked infections) from a negative binomial distribution of parameters $\kappa$ and $p = \frac{\kappa}{\kappa + R}$, where $\kappa$ is the dispersion parameter and $R$ the average number of secondary infections. (When $\kappa \to \infty$, the distribution becomes Poisson.) \item The positions of these secondary infections are then drawn such that they are located at an average distance $r_1$ from their infectors' residential locations, at any angle (isotropy). Again, this step corresponds to a conservative, unrealistic scenario in which market-linked cases infect others near their residences, and not near the market. We then repeat steps (2) and (3) for a few generations, keeping the same average distance $r_1$ for all subsequent generations of infections. We finally compute the median distance of all cases of a given generation to the market, compare it to the median distance of linked cases, and estimate the proportion of simulations for which the median distance to the market of a generation of cases is smaller than the median distance of linked cases to the market.  
\end{inparaenum} 

\begin{figure}
	\begin{tabular}{ll}
		$(a)$ $\kappa = 0.4$ & $(b)$ $\kappa = 100$ \\
		\includegraphics[width = 0.45\textwidth]{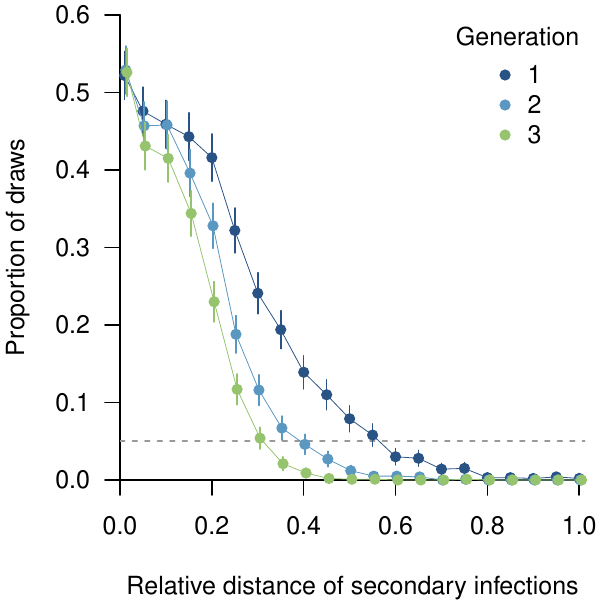}
		&
		\includegraphics[width = 0.45\textwidth]{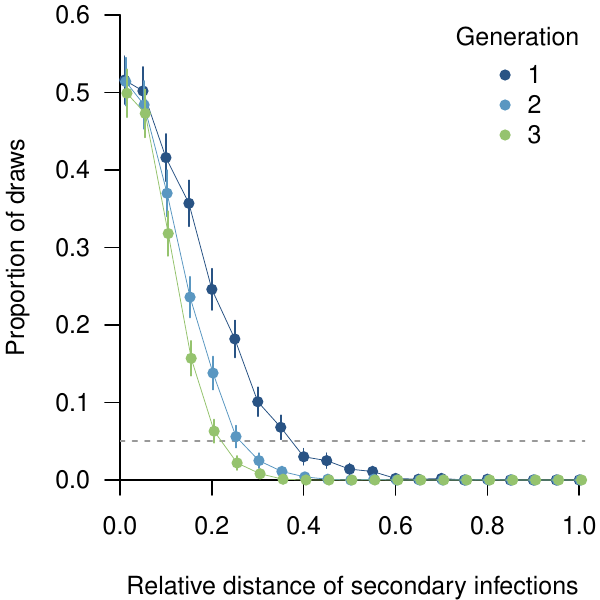}
	\end{tabular}
\caption{\textbf{Secondary infections are not necessarily further away from the market than market-linked infections.} Proportion of simulations for which the median distance to the market of infections of generation $i$ of secondary cases was smaller than the median distance to the market of market-linked cases, as a function of the relative distance of secondary infections $r_1/r_0$, for multiple generations of infections, and for a negative binomial distribution of parameter $\kappa = 0.4$ (a) or an approximately Poisson distribution (b). Other parameters: $n_0 =50$, $R=2.5$; $1000$ replicates for each point. The horizontal dashed line is at $5\%$.}
\label{fig:resultsprop}
\end{figure}

Figure~\ref{fig:resultsprop} shows that, by chance, the median distance to the market of secondary, unlinked, infections can be smaller than the median distance to the market of market-linked infections -- even under the conservative assumption that market-linked cases infected others near where they live, and not near the market. As expected, the probability that this happens gradually decreases the later the generation of infections (compare the different curves on each panel in Figure~\ref{fig:resultsprop}). Importantly, the probabilities that the median distance to the market is lower for residential locations of unlinked, versus linked, cases is higher when the mean infector-infected distance of secondary infections, $r_1$ (here assumed to be in the residential neighbourhood), is smaller than the mean distance between market-linked infections and the market, $r_0$ (including commuting distances). Finally, comparison of panels (a) and (b) also shows that unlinked cases have greater chances to be closer to the market than market-linked cases when the distribution of secondary infection numbers is over-dispersed (lower $\kappa$), as observed for \Covid{} transmission \citep{Wegehaupt2023}. The average number of secondary infections $R$ being fixed, a greater over-dispersion (lower $\kappa$) implies that the probability of an infected individual not infecting anyone is higher. Individuals at rare distant locations are therefore less likely to be included in the set of infectors when $\kappa$ is lower, and this increases the overall chance that the median distance to the origin of secondary infections is smaller than the median distance to the origin of primary infections.

Real-world data are imperfect, and the early \Covid{} case data are probably not entirely devoid of bias. Importantly, however, the case datasets were obtained retrospectively. Case definitions used in retrospective studies did not require a link to the market, since these studies occurred after that criterion was dropped from the definition \citep{Tsang2020, WHO2021, DW2024}. 

We have shown that simple mechanisms explain why the residential locations of secondary infections may be closer to the initial source than the residential locations of source-linked infections. This can happen in the absence of any detection bias, and without any constraint on the angles of infection locations, as initially described in \citet{Worobey2022}. \Citet{Weissman2024}'s premise is therefore invalidated, as is his assertion of \quotestyle{major proximity ascertainment bias} in the December 2019 COVID-19 data.

\begin{CJK*}{UTF8}{bsmi} 
To conclude, we would like to point out that actual, documented SARS-CoV-2 exposures and infections that took place early in the outbreak are especially illuminating here. At least three private clinics served both Huanan market workers and residents in the neighbourhoods surrounding the market, including one that backed onto the market itself and one on the ground floor of a residential building nearby.\footnote{特稿|与新冠病毒搏斗的民营诊所医生 (Feature: Private clinic doctors fighting against COVID-19), Caixin, April 2020, \url{https://china.caixin.com/2020-04-07/101539697.html}, archived: \url{https://archive.fo/ODg7U}} These clinics treated at least 11 of the first 27 \Covid{} patients known by Wuhan health authorities,\footnote{Caixin, \textit{ibid.}} including one of the earliest-known \Covid-patient, a shrimp vendor who worked near wildlife/meat sellers and believed she had been infected by them.\footnote{独家丨寻找华南海鲜市场第一个感染者 (Exclusive: Looking for the first infected person in Huanan Seafood Market), The Paper, March 2020,  \url{https://www.thepaper.cn/newsDetail_forward_6681256}; How It All Started: China’s Early Coronavirus Missteps, The Wall Street Journal, March 2020, \url{https:// www.wsj.com/articles/how-it-all-started-chinas-early-coronavirus-missteps-11583508932.}} 
Tragically, an owner of one of these clinics, who also lived near the market, serves as a human illustration of how the virus \quotestyle{quietly [penetrated] into} the communities surrounding the market:\footnote{Caixin, \textit{ibid.}} after treating several of the earliest patients in Wuhan -- Huanan market workers -- Dr. Liu Deyan was infected in late December 2019, and later died.\footnote{Caixin, \textit{ibid.}} \end{CJK*}

\section*{Acknowledgements} 
We thank Alex Crits-Christoph for comments and discussions.

\section*{Data availability}

The R code for the simulation is currently available at \url{https://gist.github.com/flodebarre/0af4ff60d375a322c1ede9c15bb5ed14}.

\end{document}